# Cochlear Implantation of Slim Pre-curved Arrays using Automatic Pre-operative Insertion Plans

Authors: Kareem O. Tawfik[a], M.D. and Mohammad M.R. Khan[b] (co-first author), M.S., Ankita Patro[a], MD, M.S., Miriam R. Smetak[c], M.D., M.S., David Haynes[a], M.D., Robert F. Labadie[d], M.D., Ph.D., René H. Gifford[e], Ph.D., and Jack H. Noble[a, b, e], Ph.D.

[a]Department of Otolaryngology – Head & Neck Surgery, Vanderbilt University Medical Center, Nashville, TN 37232, USA
[b]Department of Electrical & Computer Engineering, Vanderbilt University, Nashville, TN 37235, USA
[c]Department of Otolaryngology-Head & Neck Surgery, Washington University School of Medicine, St. Louis, MO 63110
[d]Department of Otolaryngology – Head & Neck Surgery, Medical University of South Carolina, Charleston, SC 29425, USA
[e]Department of Hearing and Speech Sciences, Vanderbilt University Medical Center, Nashville, TN 37232, USA

**Acknowledgements:** This work was supported in part by NIH grants R01DC008408 and R01DC014037 from the National Institute on Deafness and other Communication Disorders and UL1TR000445 from the National Center for Advancing Translational Sciences. The content is solely the responsibility of the authors and does not necessarily reflect the views of these institutes. The temporal bones used in this study were purchased with support by the Robert H. Ossoff, DMD, MD Endowed Directorship for Translational Research in Otolaryngology in the Dept. of Otolaryngology – Head and Neck Surgery at Vanderbilt University Medical Center.

**Conflicts**: KOT has served as an advisory board member for GlaxoSmithKline. DSH is a consultant for Advanced Bionics, Cochlear Americas, MED-EL GmbH, Stryker, Synthes, Grace Medical, and Oticon. RHG is a consultant for Advanced Bionics, Akouos, and Cochlear Americas, is on the clinical advisory board for Frequency Therapeutics, and is on the Board of Directors for the American Auditory Society.

**ABSTRACT**:

**Hypothesis:** Pre-operative cochlear implant (CI) electrode array (EL) insertion plans created by automated image analysis methods can improve positioning of slim pre-curved EL.

**Background:** This study represents the first evaluation of a system for patient-customized EL insertion planning for a slim pre-curved EL.

**Methods:** Twenty-one temporal bone specimens were divided into experimental and control groups and underwent cochlear implantation. For the control group, the surgeon performed a traditional insertion without an insertion plan. For the experimental group, customized insertion plans guided entry site, trajectory, curl direction, and base insertion depth. An additional 35 clinical insertions from the same surgeon were analyzed, 7 of which were conducted using the insertion plans. EL positioning was analyzed using post-operative imaging auto-segmentation techniques, allowing measurement of angular insertion depth (AID), mean modiolar distance (MMD), and scalar position.

**Results:** In the cadaveric temporal bones, 3 scalar translocations, including 2 foldovers, occurred in 14 control group insertions. In the clinical insertions, translocations occurred in 2 of 28 control cases. No translocations or folds occurred in the 7 experimental temporal bone and the 7 experimental clinical insertions. Among the non-translocated cases, overall AID and MMD were 401±41º and 0.34±0.13 mm for the control insertions. AID and MMD for the experimental insertions were 424±43º and 0.34±0.09 mm overall and were 432±19º and 0.30±0.07 mm for cases where the planned insertion depth was achieved.

**Conclusions:** Trends toward improved EL positioning within scala tympani were observed when EL insertion plans are used. Variability in MMD was significantly reduced (0.07mm vs 0.13 mm, p=0.039) when the planned depth was achieved.

**Introduction**

Cochlear implantation (CI) is a standard treatment approach for individuals experiencing moderate-to-profound sensorineural hearing loss [1]. CI electrode arrays (ELs) can be broadly divided into two categories: straight and pre-curved. Pre-curved EL are designed to approximate the curvature of the average human cochlea's modiolar wall, allowing for perimodiolar positioning [8,9]. The intra-cochlear positioning of pre-curved arrays is associated with speech recognition [10-14]. Location within the scala tympani (ST) is consistently shown to be associated with better audiologic outcomes [10,11,13,14,21]. Close proximity of the electrodes to the modiolus is hypothesized to decrease spread of excitation and has been shown to be associated with better speech recognition outcomes [10-12]. Angular insertion depth (AID) of the tip of the array measured relative to the center of the round window has been associated with better speech recognition outcome in some studies, e.g., [44], and worse speech recognition outcomes in others, e.g. [10, 11]. Pre-curved arrays are designed to achieve perimodiolar seating at a specific depth. Low AID can result in lack of stimulation in the apical portion of the cochlea and might result in extra-cochlear electrodes that do not provide auditory benefit [40]. High-AID due to over-insertion could lead to poor perimodiolar seating of the array and potential intra-cochlear trauma. Thus, a possible reason for the conflicting findings regarding AID is that both high and low AID are negatively associated with outcomes.

While associations between EL positioning and outcomes have been reported in numerous studies, styletted pre-curved EL have been associated with a high risk of scalar translocation [15] and often do not achieve the intended ideal perimodiolar positioning [16]. Variability in EL positioning is not surprising given that cochlear anatomy is variable [17-20]. Therefore, pre-curved EL insertion planning techniques based on patient-specific cochlear anatomy have been proposed to improve EL positioning and patient outcomes [22, 16].

One of the main barriers for EL insertion planning is the difficulty in accurately segmenting cochlear anatomy due to the resolution of available pre-operative images. While tremendous progress has been made in MRI imaging of the ear [23], computed tomography (CT) offers the best resolution and ability to localize key landmarks utilized by surgeons such as the chorda tympani [24]. Yet, visualization of intra-cochlear structures such as the scala tympani (ST), scala vestibuli (SV), and modiolus can be challenging with CT. Previously published auto-segmentation techniques utilize a high resolution micro-CT (µCT) non-rigid cochlear atlas [25,26, 41] to infer the position of fine structures of patients' cochleae that cannot be resolved with clinical CT with high accuracy. Other groups have developed synchrotron-based atlases but have not developed approaches to non-rigidly register them to in vivo patient CTs [45].

A novel method to create patient-specific insertion trajectories using similar segmentation techniques to identify intra-cochlear anatomy as well as the facial nerve, chorda tympani, and ossicles has been evaluated [22, 27, 28]. Previous temporal bone insertion experiments with the Advanced Bionics (Valencia, CA) Mid-Scala electrode showed statistically significant improvement in EL positioning within the ST when insertion planning was used [22]. In the current study, we compared EL positioning with and without CT-based insertion planning for a slim pre-curved EL by an experienced attending surgeon who was inexperienced with CT-based insertion planning methods, in order to evaluate whether the planning approach is effective for another EL and surgeon.

**Methods**

*Experimental protocol*

Twenty-one cadaveric temporal bone specimens were used. Each cochlea was implanted with a slim pre-curved EL (CI532/632; Cochlear Americas, Centennial, CO) by a surgeon who was experienced with this EL. The bones were evenly divided into three groups: experimental group (*Exp*), control group 1 (*C1*), and control group 2 (*C2*). To measure baseline performance in EL positioning, the *C1* group was implanted without insertion plans. Next, the *Exp* group was implanted using customized EL insertion plans designed using pre-operative cone-beam CT. Plans were created using an extended version of the previously described approach [22]. The plan included instructions on insertion trajectory (pitch, yaw, and roll) and insertion depth. Last, the *C2* group was implanted without insertion planning to detect potential changes in surgical technique following exposure to insertion planning.

Following institutional review board approval, an analysis of patients who underwent cochlear implantation with a slim pre-curved EL by the same surgeon without custom insertion plans was conducted. After removing subjects for whom there were cochlear abnormalities, such as ossification or malformations, as well as revision surgeries, 28 implanted ears from 25 patients were identified with pre- and post-operative CT scans for analysis. Additionally, 7 ears of 7 patients were implanted by the same surgeon using customized EL insertion plans in a period of time ranging from 4 months to 15 months after the completion of the temporal bone implantations. A clinical database was utilized to determine demographic information for the 32 patients as well as word recognition rates at least 6 months after implantation. Word recognition rates were assessed using the Consonant-Nucleus-Consonant (CNC) word recognition test (50-word list presented at 60 dB SPL) [42]. Test scores were collected as part of clinical routine in the unilateral, electric-only listening condition, as well as the bimodal (CI + contralateral hearing aid) condition for bimodal listeners. Of the 35 implanted ears, scores were available for 17 ears. If scores from more than one time point were available, the maximum score was recorded.

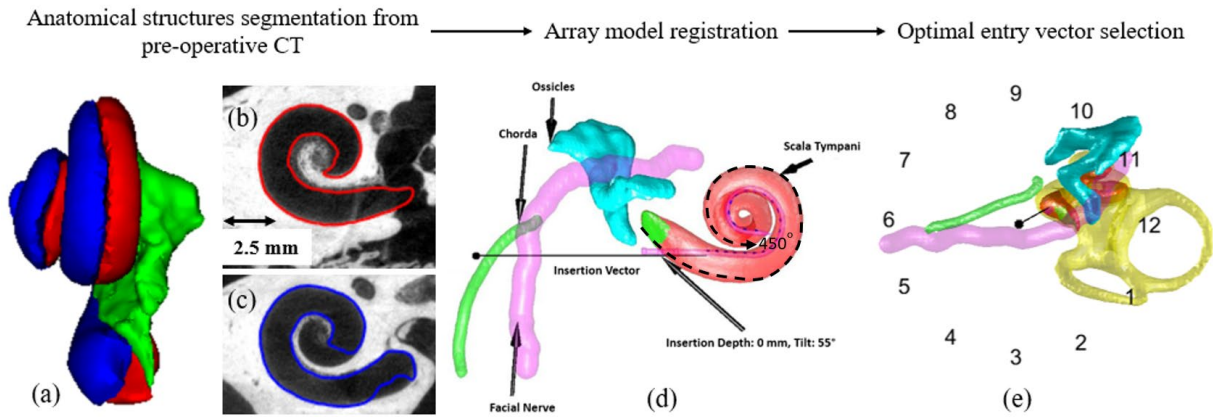

Figure 1: The steps for the proposed EL insertion planning system are shown. The intracochlear cavity atlas surface models are shown in (a). In (b) and (c) superimposed contours on one of the microCT atlases are shown. In (d), a model of the EL aligned with the medial wall of the automatically segmented ST is shown. In (e) the selected optimal trajectory (black line) for a substantially extended RW entry site is shown.

*Electrode Insertion Planning*

Previously published techniques were used to create customized EL insertion plans based on pre-operative CT imaging [22]. In brief, (1) measurements of the EL were used to create a model that represents the resting-state shape of the EL; (2) the structures of interest, including the ST, facial nerve, chorda, and ossicles, were automatically segmented in the pre-operative CT scan; (3) the array model was automatically registered, i.e, aligned, to the patient's ST modiolar wall to determine the positioning of the EL that would achieve ideal seating of the EL against the patient's modiolus (an example result is shown in Figure 1d); and (4) insertion angles (roll, pitch, yaw) and depth (referenced to the 3 white markers placed by the manufacturer at the base of the EL) were determined to match the positioning of the registered array model.

This method was modified to add an over-insertion, followed by pull-back technique [29]. The base insertion depth was defined as the distance of the middle depth marker on the array to the point where the RW plane and the insertion vector intersect (see Figure 1d). An initial over-insertion to a depth of 2mm beyond the recommended final array model base insertion depth was recommended, followed by pull-back of the array until the planned base insertion depth was achieved. This technique has been shown to achieve a more perimodiolar position of the EL compared to simply stopping at the planned depth without additional manipulation [29, 43]. Though AID was not reported in [29], we hypothesized the pull-back technique should also lead to achieving more ideal AID. The slim pre-curved array is designed to achieve good fit to the modiolus at approximately 450 degrees AID since the resting shape of the array curls approximately 450 degrees into the cochlea (as can be observed in Figure 1d).

In each case, three solutions were generated that corresponded to three different cochlear entry points: middle of the round window, a slightly extended round window, and a substantially extended round window. From the three optional plans, the surgeon selected the insertion plan that was most feasible with the given anatomy. An example of a selected insertion plan entry vector is shown in Figure 1e. In this example, the optimal trajectory, which uses a substantially extended round window and is collinear with the ST base, takes a path that is closer to the facial nerve than the chorda.

The insertion plan was presented to the surgeon in both an interactive 3D rendering (Figure 2) and a simple text format with reference to structures that can be visualized directly. The following is an example of plan instructions in text format:

> ***Entry site:*** Substantially Extended RW.
>
> ***Insertion vector:*** Distance of the insertion trajectory from the facial nerve: 1.5 mm. Distance of the insertion trajectory from the chorda: 0.5 mm. Distance of the insertion trajectory from the ossicles: 3 mm. Tilt of the optimal trajectory with the round window plane (0 degrees indicates perpendicular insertion): 55 degrees. Curl Direction (considering clock-face centered on the entry site and the stapes footplate is at 12 o'clock): 11:30. Round window insertion site (considering clock-face centered on middle of round window and the stapes footplate is at 12 o'clock): 07:30.
>
> ***Base insertion depth:*** The proximal (closest to the surgeon) marker should be inserted until it is 1.5 mm past the entry point.
>
> ***Pullback:*** After the array is inserted with the proximal (closest to the surgeon) marker 1.5 mm inside the entry point, then pullback the array until the middle marker is 0.5 mm outside the entry point.

The curl direction as well as the insertion point were presented as a clock face position. The EL insertion site was presented with a clock face centered at the RW center and using the stapes footplate as the 12:00 reference angle (Figure 2a). For determining curl direction, the center of the stapes footplate served as the 12:00 reference angle with the center of the clock face at the EL entry site (Figure 2b).

*Analysis*

Post-operative cone beam CT was used to evaluate the EL position using previously reported automated methods [30, 31]. Scalar location, mean modiolar distance (MMD), defined as the average distance from the center of each contact to the closest point on the modiolar wall, apical mean modiolar distance (AMD), defined as the average distance from the most apical 11 contacts to the closest point on the modiolar wall, and the angular insertion depth (AID) for each EL were measured. EL position was recorded twice for the experimental temporal bones, once before pullback via cone beam CT, then again after pullback via a subsequent CT.

Effectiveness of EL insertion planning in improving EL placement was determined by comparing key markers of EL positioning in the *C1* and *Exp* specimen groups. The *C2* group was used to measure whether the planning method led to changes in surgical technique. Similarly, scalar location, MMD, AMD, and AID were used to measure effectiveness of insertion plans in the clinical dataset compared to control cases where no plan was used. To analyze the impact of experimental plans when array folding or translocation does not occur, summary EL position statistics for MMD, AMD, and AID were conducted while leaving out folded and translocated cases. As the Kolmogorov-Smirnov test indicated non-normality of EL position metrics, statistically significant differences between specimen group or patient group means were assessed using Mann Whitney U-tests with a threshold of $p<0.05$. Paired t-tests with a threshold of $p<0.05$ were used to assess statistically significant differences within the experimental specimen group across the before and after pullback conditions. Brown-Forsythe variation tests with a threshold of $p<0.05$ were used to assess differences in variation across groups.

For the experimental groups, the difference between the recommended final depth of the middle marker and the actual final depth of the middle marker measured using post-implant CT, which we refer to as "base depth error" was recorded. The depth error indicates how well the planned final insertion depth was achieved. In the prior work, it was found that smaller depth errors were significantly associated with better MMD and that depth errors less than 1.5mm tended to be sufficient to lead to good placement [22]. Therefore, in addition to comparing the full experimental group to the control cases, we also perform a comparison of the sub-group of experimental cases where the absolute base depth error was less than 1.5 mm. This analysis is useful to understand how EL positioning could be improved if improvements were made to the system to provide the surgeon active feedback on EL positioning.

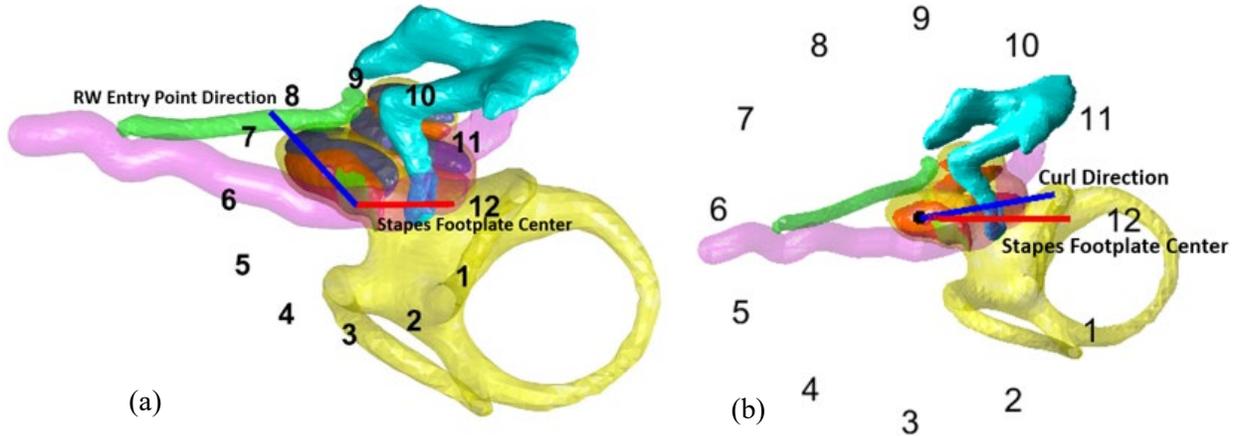

Figure 2. Shown in (a) is the clock face representation of the entry site location relative to the center of the RW. The center of the stapes footplate defines 12:00, and the center of the clock face is the RW center. Shown in (b) is the clock face representation of the array curl direction. The center of the stapes footplate defines 12:00. The center of the clock face is the EL entry site while viewing down the selected trajectory.

**Results:**

*Temporal bone study*

Individual specimen EL position analysis results for each test group are presented in Table 1(a) and summary statistics are presented in Table 1(b). In (a), each row corresponds to an individual specimen. For the experimental condition specimens, there is a separate row for before and after pullback. For the control conditions, the group (*C1* or *C2*) is listed in the group column. *D* indicates the base depth error. Negative errors indicate the actual depth is shallower than the depth suggested by the plan, and positive errors indicate the actual depth is deeper than the depth suggested by the plan. Scalar location is denoted "ST" for ELs that are entirely within the scala tympani and "ST/SV" for those for which some electrodes translocate to the scala vestibuli (SV). The column "Folded" contains "Y" for ELs that were folded in the cochlea and "N" otherwise. The AID, MMD, and AMD columns report these measurements for each EL.

In (b), counts of the number of ELs that have translocated or folded in each group are shown in the "Scalar translocations" and "Folded array" columns. The AID, MMD, and AMD columns report the mean as well as the standard deviation in parentheses for each measure for each group. Two scalar translocations were observed in C1, where one was also a folded array. One array in C2 translocated and folded. No translocations or folds were observed in the experimental group. Summary EL position statistics on cases without translocations (WT) and folds are shown in the "Control WT" rows. In the bottom of the table, p-values for statistical comparisons between groups are shown. *Exp vs Control WT* indicates all experimental cases in the *Exp* group (after pullback) compared to the entire control set (*C1 WT* and *C2 WT* combined). The remaining comparisons (*Exp* vs *C1 WT*, *C1 WT* vs *C2 WT*, and Before Pullback vs *Exp*) indicate comparisons between these specific groups. As seen in Table 1b, no statistically significant differences were detected for any of the comparisons at the current sample size.

| Specimen | Condition | Group | D (mm) | Scalar location | Folded | AID (°) | MMD (mm) | AMD (mm) |
|---|---|---|---|---|---|---|---|---|
| 1 | Control | C1 | | ST/SV | N | 336 | 0.63 | 0.61 |
| 2 | | | | ST | N | 459 | 0.22 | 0.13 |
| 3 | | | | ST | N | 435 | 0.14 | 0.13 |
| 4 | | | | ST | N | 429 | 0.31 | 0.07 |
| 5 | | | | ST | N | 430 | 0.25 | 0.12 |
| 6 | | | | ST/SV | Y | 199 | 0.83 | 0.75 |
| 7 | | | | ST | N | 387 | 0.32 | 0.22 |
| 8 | Experimental | Before Pullback | | ST | N | 428 | 0.30 | 0.07 |
| 9 | | | | ST | N | 438 | 0.39 | 0.23 |
| 10 | | | | ST | N | 443 | 0.41 | 0.23 |
| 11 | | | | ST | N | 416 | 0.47 | 0.36 |
| 12 | | | | ST | N | 357 | 0.30 | 0.18 |
| 13 | | | | ST | N | 416 | 0.33 | 0.14 |
| 14 | | | | ST | N | 434 | 0.19 | 0.16 |
| 8 | | After Pullback (Exp) | -1.20 | ST | N | 441 | 0.28 | 0.09 |
| 9 | | | -1.10 | ST | N | 412 | 0.35 | 0.12 |
| 10 | | | -1.30 | ST | N | 416 | 0.40 | 0.20 |
| 11 | | | -0.50 | ST | N | 425 | 0.34 | 0.22 |
| 12 | | | -1.70 | ST | N | 342 | 0.39 | 0.18 |
| 13 | | | -0.30 | ST | N | 409 | 0.40 | 0.16 |
| 14 | | | -0.70 | ST | N | 426 | 0.19 | 0.11 |
| 15 | Control | C2 | | ST | N | 322 | 0.48 | 0.48 |
| 16 | | | | ST | N | 431 | 0.26 | 0.40 |
| 17 | | | | ST | N | 438 | 0.18 | 0.10 |
| 18 | | | | ST/SV | Y | 188 | 1.01 | 0.88 |
| 19 | | | | ST | N | 374 | 0.44 | 0.12 |
| 20 | | | | ST | N | 425 | 0.32 | 0.20 |
| 21 | | | | ST | N | 387 | 0.37 | 0.20 |

(a)

| Condition | Group | Scalar translocations | Folded arrays | AID (degrees) | MMD (mm) | AMD (mm) |
|---|---|---|---|---|---|---|
| Control | C1 | 2 | 1 | 382 (84) | 0.38 (0.23) | 0.29 (0.25) |
| | C2 | 1 | 1 | 366 (82) | 0.44 (0.25) | 0.34 (0.26) |
| Control WT | C1 | | | 428 (23) | 0.31 (0.15) | 0.21 (0.20) |
| | C2 | | | 396 (41) | 0.34 (0.10) | 0.25 (0.14) |
| Experimental | Before pullback | 0 | 0 | 419 (27) | 0.34 (0.08) | 0.20 (0.08) |
| | Exp | 0 | 0 | 410 (30) | 0.34 (0.07) | 0.15 (0.05) |
| Mann-Whitney U-test p-value | Exp vs Control WT | | | 0.5869 | 0.2976 | 0.6507 |
| | Exp vs C1 WT | | | 0.1675 | 0.0618 | 0.6847 |
| | C1 WT vs C2 WT | | | 0.2353 | 0.1207 | 0.3153 |
| Paired t-test p-value | Before Pullback vs Exp | | | 0.3151 | 0.4757 | 0.1152 |

(b)

Table 1. Electrode position analysis results for each specimen the temporal bone study (a) and summary statistics for each condition and group (b).

*Clinical study*

Individual EL position analysis results for patients in the clinical database are presented in Table 2(a). The first 11 columns indicate the subject number (Sub.), ordered by date of surgery from earliest to most recent; the condition ("Cont" indicates control and "**Exp**" indicates a custom insertion plan was used); the laterality ("Side"), where "L" indicates left ear and "R" indicates a right ear; the duration of hearing loss in years prior to implantation ("Dur HL"); the wear time in hours per day as measured using datalogging; age at implantation in years; age at testing in years; everyday hearing configuration, which was either unilateral implant ("Unil"), bilateral implant ("Bilat"), or bimodal ("Bimod"); whether the ear is the first implanted ear ("1$^{st}$ Ear"), with "N" indicating it was not the patients first implanted ear and "Y" indicating it was; Etiology of hearing loss, which was either unknown or due to noise, meningitis, Meneire's disease, Otosclerosis ("Otoscler."), familial hearing loss, Auditory neuropathy spectrum disorder ("ANSD"), or Neurofibromatosis Type II ("NF2"); and biological sex, with "F" indicating female and "M" indicating male. The next 5 columns correspond to EL position metrics and are defined identically to Table 1. A column indicating folding was not included as there were no folded arrays in this dataset. The final two columns contain CNC word recognition scores recorded in the unilateral, implant only condition and in the bimodal condition, when available.

Summary statistics on EL positioning and word recognition rates are presented in Table 2(b). In the bottom of the table, p-values for statistical comparisons between groups are shown. *"Cont. WT vs Exp. All"* indicates all 26 non-translocated cases in the control group compared to the 7 cases in the entire experimental group. "Before vs Exp. All" indicates a comparison between the 6 control cases implanted before starting the clinical study and the entire experimental group. "Before vs After WT' indicates a comparison between the "Before" and "After" control sub-groups without the translocations. No statistically significant difference in average EL positioning was detected with the current sample size.

The first case in the experimental group, subject 6 (S6) can be observed to be deeply over-inserted to 542 degrees (see Table 2a). The base insertion depth error in this case was 2.7 mm, indicating the base of the array was inserted substantially past the planned depth, which likely led to the high AID, MMD, and AMD values. S12 had a base insertion depth error of -2.2 mm, indicating it is much shallower than the planned depth. The experimental $|D|<1.5$ mm sub-group reports EL summary statistics when not including these two cases where the actual depth did not well match the planned depth.

In Table 3, results across the temporal bone and clinical cases are pooled, showing average and standard deviation statistics on AID, MMD, and AMD for the 37 Control WT cases, 14 experimental cases, and the 11 experimental cases where the absolute base depth error was less than 1.5 mm. In the bottom of the table, p-values are shown for Mann Whitney U-tests and Brown Forsythe variation tests for significant differences between the control and two experimental groups. No significant differences in rank are detected by the Mann Whitney U-test at this sample size. The Brown Forsythe test did not detect significance for MMD or AMD, but revealed a significant difference in standard deviation of MMD between the Control WT and the experimental group with depth error less than 1.5 mm.

Scatter plots with correlation regression lines are shown in Figure 3 comparing AID error in degrees (assuming an ideal depth of 450 degrees) with CNC word scores in (a), MMD with CNC word scores in (b), and AMD with CNC word scores in (c) across all 17 subjects for whom word scores were available. The Pearson correlation coefficient between CNC word scores versus AMD and MMD and was found to not be significant at this sample size, with R= -0.34 (p=0.19) and R= -0.37 (p=0.14). The correlation between AID depth error and CNC word scores was found to be significant with R= -0.53 (p=0.03).

| Sub. | Cond ition | Side | Dur. HL (Yrs) | Wear time (hrs/day) | Age at implant-ation (Yrs) | Age at Testing (Yrs) | Hearing Configur ation | 1st Ear | Etiology | Sex | Base depth error (mm) | Scalar loca-tion | AID (degrees) | MMD (mm) | AMD (mm) | CNC Words - implant only (%) | CNC Words - bimodal (%) |
|---|---|---|---|---|---|---|---|---|---|---|---|---|---|---|---|---|---|
| 1 | Cont | L | 8 | 7.7 | 42 | 43 | Bilat | N | ANSD | F | | ST | 331 | 0.15 | 0.20 | 50 | |
| 2 | Cont | R | | 12.6 | 79 | 80 | Unil | Y | Noise | M | | ST | 395 | 0.43 | 0.19 | 74 | |
| 3 | Cont | L | | | 56 | | Bimod | Y | Unknown | F | | ST | 384 | 0.27 | 0.11 | | |
| 4 | Cont | L | | | 1 | | Bilat | Y | Unknown | F | | ST | 430 | 0.50 | 0.23 | | |
| 4 | Cont | R | | | 1 | | Bilat | Y | Unknown | F | | ST | 414 | 0.36 | 0.17 | | |
| 5 | Cont | R | | | 3 | | Unil | Y | Unknown | F | | ST | 376 | 0.43 | 0.09 | | |
| 6 | **Exp** | L | 27 | 15.8 | 38 | 39 | Bilat | N | Unknown | F | 2.7 | ST | 542 | 0.58 | 0.57 | 22 | |
| 7 | **Exp** | L | | | 21 | | Unil | Y | Meningitis | F | -0.5 | ST | 451 | 0.28 | 0.22 | | |
| 8 | **Exp** | L | 1 | 14 | 72 | 73 | Bimod | Y | Noise | F | -0.8 | ST | 445 | 0.32 | 0.17 | 80 | 80 |
| 9 | Cont | L | | 10.2 | 58 | 60 | Unil | Y | Meniere's | F | | ST | 377 | 0.50 | 0.33 | 58 | |
| 10 | **Exp** | R | 13 | 8.1 | 18 | 19 | Bimod | Y | Unknown | F | 0.2 | ST | 391 | 0.21 | 0.25 | 86 | 86 |
| 11 | **Exp** | R | 9 | 14.2 | 60 | 61 | Bimod | N | Noise | M | -0.7 | ST | 423 | 0.26 | 0.17 | 74 | 82 |
| 12 | **Exp** | R | | 12.5 | 40 | | Bilat | Y | Meningitis | F | -2.2 | ST | 407 | 0.40 | 0.07 | | |
| 13 | **Exp** | L | 22 | 16.6 | 59 | 61 | Bimod | Y | Meniere's | M | 0.1 | ST | 401 | 0.23 | 0.30 | 88 | 90 |
| 14 | Cont | R | | | 4 | | Bilat | Y | Unknown | M | | ST | 411 | 0.17 | 0.17 | | |
| 15 | Cont | L | | 13.7 | 82 | 83 | Unil | Y | Noise | M | | ST | 444 | 0.22 | 0.12 | 70 | |
| 16 | Cont | L | | 15 | 28 | 29 | Bimod | Y | Otoscler. | M | | ST | 318 | 0.36 | 0.16 | 30 | 72 |
| 17 | Cont | L | | | 4 | | Unil | Y | Unknown | M | | ST | 415 | 0.55 | 0.61 | | |
| 18 | Cont | L | 31 | 12.1 | 36 | 36 | Unil | Y | Unknown | F | | ST | 363 | 0.37 | 0.10 | 84 | |
| 19 | Cont | L | | 14.5 | 79 | 79 | Unil | Y | Noise | M | | ST | 407 | 0.22 | 0.08 | 60 | |
| 20 | Cont | L | 20 | 11.5 | 69 | 69 | Bimod | Y | Familial | F | | ST | 442 | 0.46 | 0.49 | 84 | 80 |
| 21 | Cont | R | | | 32 | | Unil | Y | NT2 | F | | ST/SV | 391 | 0.66 | 0.67 | | |
| 22 | Cont | L | | 7.8 | 19 | 19 | Unil | Y | Unknown | F | | ST | 388 | 0.29 | 0.29 | 34 | |
| 23 | Cont | R | 66 | 22.5 | 70 | 71 | Bilat | N | Unknown | F | | ST | 348 | 0.43 | 0.33 | 42 | |
| 24 | Cont | L | | | 6 | | Bilat | Y | Unknown | M | | ST | 433 | 0.13 | 0.12 | | |
| 24 | Cont | R | | | 6 | | Bilat | Y | Unknown | M | | ST | 422 | 0.43 | 0.13 | | |
| 14 | Cont | L | | | 5 | | Bilat | N | Unknown | M | | ST | 399 | 0.14 | 0.18 | | |
| 25 | Cont | R | 11 | 11.2 | 61 | 62 | Bimod | Y | Noise | M | | ST | 446 | 0.38 | 0.26 | 52 | 94 |
| 26 | Cont | L | | | 46 | | Bimod | Y | Unknown | M | | ST | 389 | 0.38 | 0.14 | | |
| 27 | Cont | L | 55 | | 73 | | Bimod | Y | Noise | M | | ST | 447 | 0.44 | 0.40 | | |
| 28 | Cont | R | | | 85 | | Bimod | Y | Noise | M | | ST | 408 | 0.31 | 0.08 | | |
| 29 | Cont | R | | | 70 | | Bimod | Y | Unknown | F | | ST | 438 | 0.17 | 0.08 | | |
| 30 | Cont | R | | | 45 | | Bimod | Y | Meniere's | F | | ST | 276 | 0.65 | 1.03 | | |
| 31 | Cont | R | 15 | 12.5 | 79 | 80 | Bimod | Y | Noise | M | | ST | 403 | 0.47 | 0.38 | 42 | 78 |
| 32 | Cont | L | | | 61 | | Bimod | Y | Meniere's | M | | ST/SV | 290 | 0.67 | 0.47 | | |

(a)

| Group | Sub-group | Scalar trans-locations | AID (degrees) | MMD (mm) | AMD (mm) | CNC Word Recog. Rate - implant only (%) | CNC Word Recog. Rate - bimodal (%) |
|---|---|---|---|---|---|---|---|
| Cont. | All (N=28) | 2 | 392 (45) | 0.38 (0.15) | 0.27 (0.22) | 54.3 (18.76) N=12 | 81.0 (8.06) N=4 |
| | Before (N=6) | 0 | 388 (31) | 0.36 (0.12) | 0.16 (0.05) | 50.0 (19.60) N=2 | |
| | After (N=21) | 2 | 394 (49) | 0.38 (0.16) | 0.30 (0.24) | 55.3 (19.28) N=9 | 81.0 (8.06) N=4 |
| | After WT (N=19) | | 400 (45) | 0.35 (0.14) | 0.27 (0.23) | | |
| Exp. | All (N=7) | 0 | 437 (48) | 0.33 (0.12) | 0.25 (0.15) | 70.0 (24.49) N=5 | 84.5 (3.84) N=4 |
| | \|D\|<1.5 (N=5) | 0 | 422 (23) | 0.26 (0.04) | 0.22 (0.05) | 82.0 (5.48) N=4 | |
| Mann-Whitney U-test p-values | Cont. WT vs Exp. All | | 0.0820 | 0.5522 | 0.6919 | | |
| | Before vs Exp. All | | 0.0633 | 0.4751 | 0.3173 | | |
| | Before vs After WT | | 0.2794 | 0.7746 | 0.6332 | | |

(b)

Table 2. Electrode position analysis results for each subject (a) and summary statistics for each condition and group (b).

| Condition | | AID (°) | MMD (mm) | AMD (mm) |
|---|---|---|---|---|
| Control WT | | 401 (41) | 0.34 (0.13) | 0.23 (0.19) |
| Exp. All (after pullback) | | 424 (43) | 0.34 (0.09) | 0.20 (0.12) |
| Exp. $|D|<1.5$mm | | 432 (19) | 0.30 (0.07) | 0.18 (0.06) |
| Mann-Whitney U-test | Control WT vs Exp. All | 0.184 | 0.792 | 0.933 |
| | Control WT vs Exp. $|D|<1.5$ | 0.141 | 0.315 | 0.932 |
| Brown-Forsythe test | Control WT vs Exp. All | 0.610 | 0.181 | 0.352 |
| | Control WT vs Exp. $|D|<1.5$ | 0.051 | **0.039** | 0.165 |

Table 3. Pooled EL position statistics across the temporal bone and clinical control WT and experimental groups. Significant p-values are in bold.

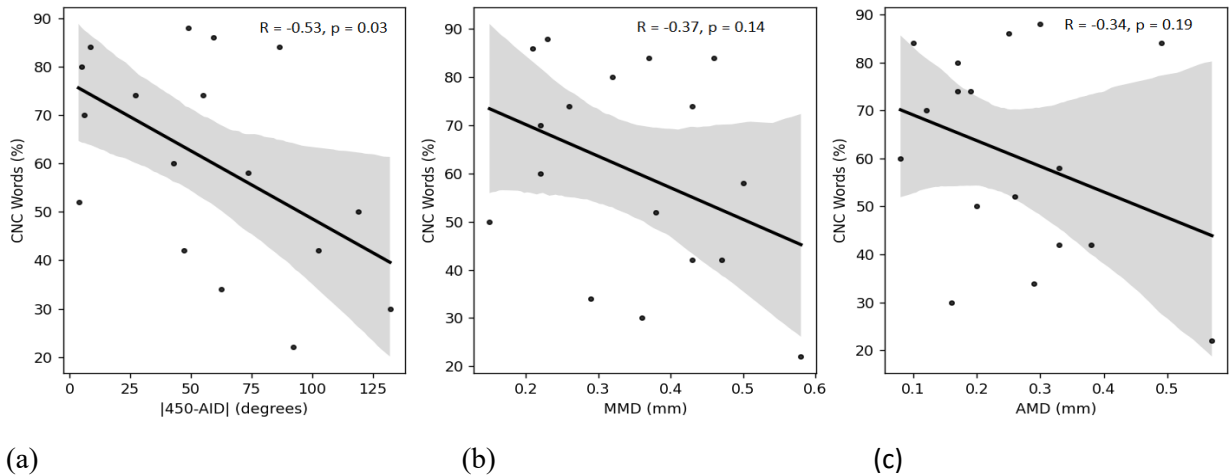

(a)  (b)  (c)

Figure 3. Scatter plots of AID error on horizontal axis in (a), MMD on horizontal axis in (b), and AMD on horizontal axis in (c) versus CNC word scores in the electric only condition on the vertical axis. A regression line (black line) and 95% confidence interval (shaded area) are also shown in each plot.

**Discussion**

      Precision medicine has opened a new dimension in the health care system. However, due to the traditional one-electrode-size-fits-most approach to CI, patient-specific EL insertion planning has yet to come into widespread use. Currently, the average postoperative sentence and word recognition are reported to be 70% and 55% [33], respectively. Customized EL insertion planning offers the opportunity to potentially improve post-operative hearing performance with improved EL positioning [11]. Indeed, even in this relatively small study, a significant correlation was found between AID error and CNC word scores (R= -0.53, p=0.03), as well as a trend towards correlation between MMD, AMD and CNC word scores. These data support that techniques that reduce AID errors, AMD, and MMD might improve hearing outcomes.

      Herein, we report the results of using customized EL insertion plans, which aim to improve EL positioning, first in a temporal bone study, then in a clinical study. While we currently are unable to measure how well the surgeon adhered to the planned insertion angles, we can measure the difference, *D*, between the planned and the actual depth of the base of the array relative to the round window measured from post-insertion CT. *D* is reported for each experimental case in Tables 1 and 2. In temporal bone 12, the actual depth was 1.7 mm shallower than the planned depth, and the resulting AID was 342 degrees, the shallowest insertion in the experimental group. In clinical case 6, the first experimental case, we observe the base depth error to be 2.7 mm, corresponding to an over-insertion and the deepest observed AID of 542 degrees, well beyond the ideal 450 degrees for this EL. The base insertion depth for case 12 was 2.2 mm shallower than the planned depth, leading to increased MMD in the most basal contacts of the array. Large base depth errors can impact overall EL positioning, as observed in both this study and the previous one conducted with a different EL [22]. Thus, separate analyses were performed for all experimental cases versus those where base depth was within 1.5 mm of the planned depth.

      In the temporal bone study, we found 0/7 experimental insertions folded or translocated. This is in contrast with the control insertions, for which 3/14 translocated (including 2 folded arrays). Array folding may lead to intracochlear trauma and worsened performance, often leading to the need for revision surgery [34]. However, folding and translocation both tend to occur more frequently in temporal bone insertions compared to clinical ones, likely due to degraded tissue integrity for ex vivo temporal bones [38]. In the non-guided condition, we indeed observed higher translocation rates in the temporal bones than reported with the clinical use of this same device

[35]. Because translocation impacts AID, MMD, and AMD, and the clinical translocation rate with this electrode is relatively low (2/28 in the control group), we also compared EL position across groups after removing the folded and translocated array cases from the control group. While no significant differences were identified with the Mann Whitney U-test in this relatively small study, there were trends toward lower AMD, MMD, and AID error. With pre-operative planning, EL positioning within ST was observed in all cases.

We include analysis of the experimental with $|D|$<1.5 mm group to observe EL positioning when the achieved depth is close to the plan. Given the current mean and standard deviation of the control WT and experimental with $|D|$<1.5 mm groups, power analysis suggests that 14 experimental cases would be needed to have a power of 0.80 to detect a significant difference in AID at the $p<0.05$ level with the Mann-Whitney U-test. Sixty-six cases and 81 cases would be needed for similar power to detect differences in MMD and AMD. We found significantly reduced variability in MMD using the Brown-Forsythe test, and a trend toward reduced variability in AID and AMD for the experimental with $|D|$<1.5 mm group compared to controls. Reduced variability suggests that if the plan is successfully followed, EL positioning can become more standardized.

Using the pull-back technique, AMD was reduced without significant change in AID. Achieving reduced modiolar distance at the apical portion of the array is of interest because channel interaction due to modiolar distance has been shown to have greater effect for apically located electrodes compared to basal electrodes [32].

We hypothesized that the use of the insertion planning approach may lead to learning effects for surgeons after using insertion planning. If substantial learning effects existed, then after using the planning tool for a series of cases, future EL placements should improve even without the use of pre-operative planning. We observed no significant differences in the pre- or post-planning insertions. Further experiments are necessary to evaluate potential learning effects.

There are several limitations to this study. First, with limited sample size, there was limited statistical power. Also, there was limited ability to measure adherence to the insertion plan. While it was possible to measure the final base insertion depth, it was not possible to measure the implemented insertion angles. This may be important as the previous study found that deviations intentionally introduced in the planned yaw angle were associated with scalar translocations and increased MMD, and deviated pitch angles were associated with increased AID error [22]. Next, only one type of EL was evaluated in this study, although the method has also been validated with another pre-curved EL [22]. Future work will be aimed at developing better strategies for communicating the plans to the surgeon as well as strategies for actively monitoring and providing feedback on the insertion, since final EL position is sensitive to successful plan implementation. Future experiments will also include multiple surgeons and increased sample size for both EL types. Although pre-curved ELs were the focus of these studies, the insertion planning approach could be applicable to straight ELs, with some modifications to the strategy for determining optimal vector and depth [36, 37].

Of the 7 patients for whom insertion plans were used clinically, CNC word recognition test scores were available for 5 cases from at least 6 months post-implantation and were found to be 70.0±24.49 %. If limiting to the 4 cases with base depth error less than 1.5mm, the scores were 82.0±5.48 %. A larger sample size is necessary to confirm using insertion plans improves hearing outcomes, however, these results are consistent with multiple studies that have shown well positioned pre-curved arrays to be associated with significantly higher speech recognition performance [10,11,13,14,40].

Cochlear Implantation using Automatic Pre-operative Insertion Plans